\documentclass[submission,copyright,creativecommons]{eptcs}
\providecommand{\event}{Festschrift Symposium for Dave Schmidt} 

\usepackage[T1]{fontenc}
\usepackage{graphicx}
\usepackage{amsmath}
\usepackage{yfonts}
\usepackage{galois}
\usepackage{amssymb}
\usepackage{listings}
\usepackage{stmaryrd}

\newcommand{\naturals}{\ensuremath{\mathbb{N}}}

\newcommand{\statement}[1]{\ensuremath{\mathsf{#1}}}
\newcommand{\funzione}[2]{\ensuremath{#1 \rightarrow#2}}
\newcommand{\pair}[2]{\ensuremath{#1 \times #2}}

\newcommand{\sem}[1]{\ensuremath{\llbracket \mathtt{#1} \rrbracket}}
\newcommand{\semanticanome}[1]{\ensuremath{\mathbb{#1}}}

\newcommand{\cset}[1]{\ensuremath{\mathsf{#1}}}

\newcommand{\cel}[1]{\ensuremath{\mathsf{#1}}}

\newcommand{\aset}[1]{\cset{\overline{#1}}}
\newcommand{\ael}[1]{\cel{\overline{#1}}}

\newcommand{\firstdomain}{\aset{A}}
\newcommand{\seconddomain}{\aset{B}}
\newcommand{\concretedomain}{\cset{C}}
\newcommand{\cartesiandomain}{\textswab{C}}
\newcommand{\powerdomain}{\textswab{P}}

\newcommand{\firstsemantics}{\semanticanome{S}_\firstdomain}
\newcommand{\secondsemantics}{\semanticanome{S}_\seconddomain}
\newcommand{\concretesemantics}{\semanticanome{S}_\concretedomain}
\newcommand{\cartesiansemantics}{\semanticanome{S}_\cartesiandomain}

\newcommand{\firstsemanticsapplication}[1]{\firstsemantics\sem{#1}}
\newcommand{\secondsemanticsapplication}[1]{\secondsemantics\sem{#1}}
\newcommand{\concretesemanticsapplication}[1]{\concretesemantics\sem{#1}}
\newcommand{\cartesiansemanticsapplication}[1]{\cartesiansemantics\sem{#1}}

\title{A Survey on Product Operators in Abstract Interpretation}
\author{Agostino Cortesi
\institute{Ca' Foscari University\\ Venice, Italy}
\email{cortesi@dsi.unive.it}
\and
Giulia Costantini
\institute{Ca' Foscari University\\ Venice, Italy}
\email{costantini@dsi.unive.it}
\and
Pietro Ferrara
\institute{ETH \\Zurich, Switzerland}
\email{pietro.ferrara@inf.ethz.ch}
}
\def\titlerunning{Product Operators in Abstract Interpretation}
\def\authorrunning{A. Cortesi, G. Costantini \& P. Ferrara}
\begin{document}
\maketitle

\begin{abstract}
The aim of this paper is to provide a general overview of the product operators introduced in the literature as a tool to enhance the analysis accuracy in the Abstract Interpretation framework.  In particular we focus on the Cartesian and reduced products, as well as on the reduced cardinal power, an under-used technique whose features deserve to be stressed for their potential impact in practical applications. 
\end{abstract}

\section{Introduction}
\begin{center}
\begin{figure}
\includegraphics[scale=0.8]{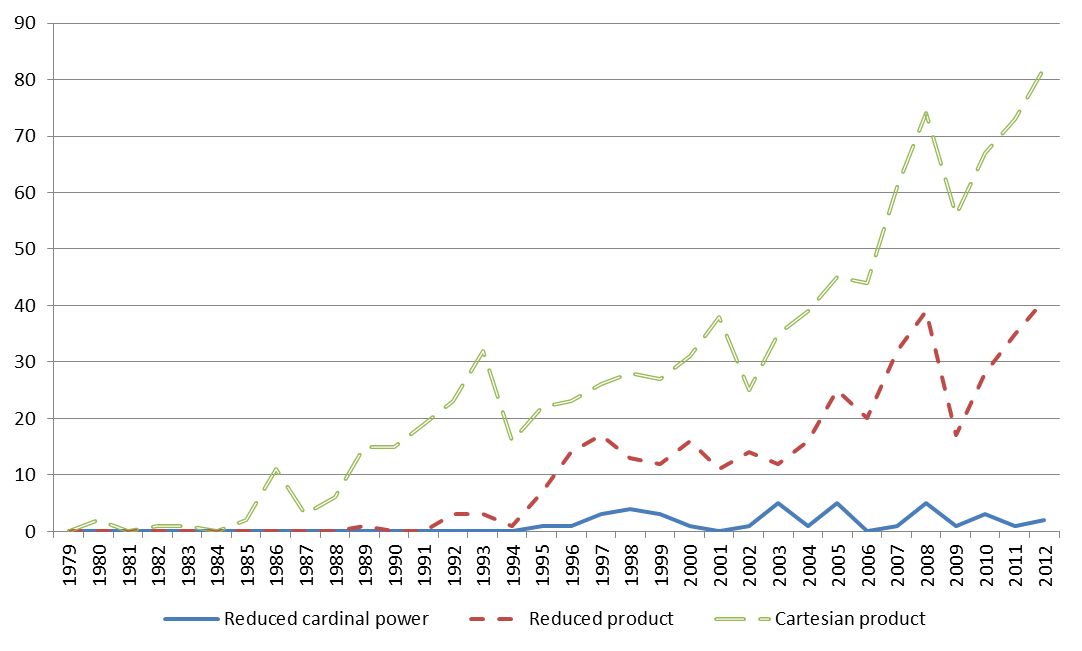}
\label{fig:citations}
\caption{Number of annual citations since 1979, for the searches "reduced cardinal power", "reduced product" and "cartesian product" (in the abstract interpretation field)}
\end{figure}
\end{center}

Abstract interpretation \cite{CC79} has been widely applied as a general technique for the sound approximation of the semantics of computer programs. In particular, abstract domains (to represent data) and semantics (to represent data operations) approximate the concrete computation. When analyzing a program and trying to prove some property on it, the quality of the result is determined by the abstract domain choice. There is always a trade-off between accuracy and efficiency of the analysis. During the years, various abstract domains have been developed. An interesting feature of the abstract interpretation theory is the possibility to combine different domains in the same analysis. In fact, the abstract interpretation framework offers some standard ways to compose abstract domains, ensuring the preservation of the theoretical properties needed to guarantee the soundness of the analysis. These compositional methods are called \emph{domain refinements}. A systematic treatment of abstract domain refinements has been given in \cite{FGR96, GR97}, where a generic refinement is defined to be a lower closure operator on the lattice of abstract interpretations of a given concrete domain. These kinds of operators on abstract domains provide high-level facilities to tune a program analysis in terms of accuracy and cost. Two of the most well-known domain refinements are the \emph{disjunctive completion} \cite{CC79, CC94, FR99, GR98, J97} and the \emph{reduced product} \cite{CC79}, but they are not the only ones. The reduced product can be seen as the most precise refinement of the simple Cartesian product. Moreover, the reduced cardinal power is introduced by \cite{CC79}. While the other domain refinements have been, since their introduction, widely used and explored, the reduced cardinal power has seen definitely less further developments since 1979, with the exception of \cite{GR99}. To verify our assertion, we looked for the number of scientific citations (in the abstract interpretation context) to some domain refinements in Google Scholar. We depicted the results of this search in Figure \ref{fig:citations}. In particular, we focused on the Cartesian product, the reduced product and the reduced cardinal power. Throughout the years, the number of citations increases in all three cases, but the absolute numbers are very different: just consider that the total citations of \textquotedblleft Cartesian product\textquotedblright\ are 964, while the ones to \textquotedblleft reduced cardinal power\textquotedblright\ are only 38. 

However, we think that the reduced cardinal power refinement has a potential which has not been completely exploited yet. Consider, for example, that such refinement is used by ASTR\'EE\cite{CCFMMMR05}, the well-known static program analyzer that proves the absence of run-time errors in safety-critical embedded applications written or automatically generated in C. This paper aims at giving a survey of the different product operators introduced in the literature by providing a uniform terminology, an analysis of their complexity, and of the implementation effort they require.

\section{Formal definitions, complexity, and implementation}
\label{sect:formaldef}
In this Section, we introduce the three main ways of combining various abstract domains in the abstract interpretation theory (namely, the Cartesian product, the reduced product, and the reduced cardinal power). For the sake of simplicity, we will focus on the combination of \emph{two} abstract domains. Therefore, we suppose that two abstract domains $\firstdomain$ and $\seconddomain$ are given, and that they are equipped with lattice operators: $\langle \firstdomain, \leq_\firstdomain, \sqcup_\firstdomain, \sqcap_\firstdomain \rangle$ and $\langle \seconddomain, \leq_\seconddomain, \sqcup_\seconddomain, \sqcap_\seconddomain \rangle$.

In addition, let $\concretedomain$ be the concrete domain. We suppose that this domain is equipped with lattice operators as well: $\langle \concretedomain, \leq_\concretedomain, \sqcup_\concretedomain, \sqcap_\concretedomain \rangle$. We suppose that both $\firstdomain$ and $\seconddomain$ are sound abstractions of $\concretedomain$, that is, they form a Galois connection: $\langle \concretedomain, \leq_\concretedomain \rangle \galois{\alpha_\firstdomain}{\gamma_\firstdomain} \langle \firstdomain, \leq_\firstdomain \rangle$ and $\langle \concretedomain, \leq_\concretedomain \rangle \galois{\alpha_\seconddomain}{\gamma_\seconddomain} \langle \seconddomain, \leq_\seconddomain \rangle$, where $\alpha_\firstdomain, \gamma_\firstdomain$ and $\alpha_\seconddomain, \gamma_\seconddomain$ are the abstraction and concretization functions of $\firstdomain$ and $\seconddomain$, respectively. \footnote{There exist other approaches which can be used as well (e.g., when the best abstraction function does not exist \cite{CH78}). However, the Galois connection-based approach is definitely the most commonly used \cite{CC92a}.}

Finally, abstract domains provide abstract semantic transformers. Formally, we suppose that $\firstdomain$ provides $\firstsemantics : \funzione{\firstdomain}{\firstdomain}$, and $\seconddomain$ provides $\secondsemantics : \funzione{\seconddomain}{\seconddomain}$. These are sound approximation of the concrete semantics $\concretesemantics : \funzione{\concretedomain}{\concretedomain}$. Formally, $\forall \ael{a} \in \firstdomain : \concretesemanticsapplication{\gamma_\firstdomain(\ael{a})} \leq_\concretedomain \gamma_\firstdomain(\firstsemanticsapplication{\ael{a}})$ and $\forall \ael{b} \in \seconddomain : \concretesemanticsapplication{\gamma_\seconddomain(\ael{b})} \leq_\concretedomain \gamma_\seconddomain(\secondsemanticsapplication{\ael{b}})$.


\subsection{Cartesian product}
The elements of this domain are elements in the Cartesian product of the two domains, and the operators are defined as the component-wise application of the operators of the two domains. 

Formally, let $\cartesiandomain=\pair{\firstdomain}{\seconddomain}$ be the Cartesian product. The partial order is defined as the conjunction of the partial orders of the two domains ($(\ael{a}_1, \ael{b}_1) \leq_\cartesiandomain (\ael{a}_2, \ael{b}_2) \Leftrightarrow \ael{a}_1 \leq_\firstdomain \ael{a}_2 \wedge \ael{b}_1 \leq_\seconddomain \ael{b}_2$). Similarly, the least upper bound and the greatest lower bound operators are defined as the component-wise application of the operators of the two domains ($(\ael{a}_1, \ael{b}_1) \sqcup_\cartesiandomain (\ael{a}_2, \ael{b}_2) = (\ael{a}_1 \sqcup_\firstdomain \ael{a}_2, \ael{b}_1 \sqcup_\seconddomain \ael{b}_2)$ and $(\ael{a}_1, \ael{b}_1) \sqcap_\cartesiandomain (\ael{a}_2, \ael{b}_2) = (\ael{a}_1 \sqcap_\firstdomain \ael{a}_2, \ael{b}_1 \sqcap_\seconddomain \ael{b}_2)$, respectively). This way, we obtain that the Cartesian product $\langle \cartesiandomain, \leq_\cartesiandomain, \sqcup_\cartesiandomain, \sqcap_\cartesiandomain \rangle$ forms a lattice. The pairwise approach to combine operators holds also in the case of widening. In fact, given  the widening operators $\nabla_A$ and $\nabla_B$ on the domains $A$ and $B$, respectively, the operator $\nabla_{A \times B}((a_1,b_1),(a_2,b_2))=(a_1 \nabla_A a_2, b_1 \nabla_B b_2)$ is a widening operator on $\cartesiandomain$ \cite{ZanioliCortesi}.

In addition, the abstraction function $\alpha_\cartesiandomain$ consists in the component-wise application of the abstraction functions of the two domains ($\alpha_\cartesiandomain(\cel{c}) = (\alpha_\firstdomain(\cel{c}), \alpha_\seconddomain(\cel{c}))$), while the concretization function $\gamma_\cartesiandomain$ consists in the intersection of the results obtained by the concretization functions of the two domains on the corresponding component ($\gamma_\cartesiandomain(\ael{a}, \ael{b}) = \gamma_\firstdomain(\ael{a}) \sqcap_\concretedomain \gamma_\seconddomain(\ael{b})$). Then, the Cartesian product forms a Galois connection with the concrete domain (formally, $\langle \concretedomain, \leq_\concretedomain \rangle \galois{\alpha_\cartesiandomain}{\gamma_\cartesiandomain} \langle \cartesiandomain, \leq_\cartesiandomain \rangle$).

Finally, also the semantic operator $\cartesiansemantics : \funzione{\cartesiandomain}{\cartesiandomain}$ is defined as the component-wise application of the abstract semantics of the two domains (formally, $\cartesiansemanticsapplication{(\ael{a}, \ael{b})} = (\firstsemanticsapplication{\ael{a}}, \secondsemanticsapplication{\ael{b}})$). This way, the semantics of the Cartesian product is a sound over-approximation of the concrete semantics ($\forall (\ael{a}, \ael{b}) \in \cartesiandomain : \concretesemanticsapplication{\gamma_\cartesiandomain(\ael{a}, \ael{b})} \leq_\concretedomain \gamma_\cartesiandomain(\cartesiansemanticsapplication{\ael{a}, \ael{b}})$).

As pointed out by Patrick Cousot \cite{MIT}, \textquotedblleft the Cartesian product discovers in one shot the information found separately by the component analyses\textquotedblright, but \textquotedblleft we do not learn more by performing all analyses simultaneously than by performing them one after another and finally taking their conjunctions\textquotedblright.

In addition, the Cartesian product may contain several abstract elements that represent the same information. For instance, consider the Cartesian product of the Interval and the Parity domains, and in particular the elements $([2..4], \cel{O})$, $([2..3], \cel{O})$, $([3..4], \cel{O})$, and $([3..3], \cel{O})$, where $\cel{O}$ represents the odd element of the Parity domain. All these elements concretize to the singleton $\{3\}$, but some of them are not minimal\footnote{An abstract element $\ael{a}$ is minimal w.r.t. a property $\cel{c} \in \concretedomain$ if and only if (i) $\gamma(\ael{a}) \geq_\concretedomain \cel{c}$ and (ii) $\not \exists \ael{a}' : \gamma(\ael{a}') \geq_\concretedomain \cel{c} \wedge \ael{a}' < \ael{a}$.}.



\subsubsection{Complexity}
When applying lattice or semantic operators, the complexity of the operator defined on $\cartesiandomain$ is exactly the sum of the complexity of the corresponding operators on $\firstdomain$ and $\seconddomain$. Instead, the height of the lattice of $\cartesiandomain$ (that is important to estimate the complexity of computing a fixpoint using this domain) is the multiplication of the heights of $\firstdomain$ and $\seconddomain$.

\subsubsection{Implementation}
Given the implementations of $\firstdomain$ and $\seconddomain$, the implementation of $\cartesiandomain$ is completely straightforward, and it could be used to combine any existing abstract domain in a completely generic way. In fact, the implementation only requires the existence of the operators, and there is no need to develop anything specific on such domains.

\subsection{Reduced product}
Even if the Cartesian product is a quite effective way to cheaply combine two domains in terms of both formalization and implementation, it is clear that one may want to let the information flow among the two domains to mutually refine them. Already in one of the foundative papers of abstract interpretation \cite{CC79}, Patrick and Radhia Cousot introduced the reduced product exactly with the purpose of refining the information tracked by $\firstdomain$ and $\seconddomain$. In particular, when we have an abstract state that is non-minimal, we can take the smallest element which represents the same information by \emph{reducing} it. A reduction improves the precision of the abstract representation
with respect to the order in the Cartesian product without affecting its concrete meaning. Intuitively, a reduction exploits the information tracked by one of the two domains involved in the product to refine the information tracked by the other one (and viceversa). Let $(a,b)$ be an element of a reduced product (where $a$ and $b$ belong respectively to the two domains combined in the product: $a \in \firstdomain$, $b \in \seconddomain$). Let $c_1$ be the set of concrete values associated to $a$ and $c_2$ be the set of concrete values associated to $b$. Then, the element $(a,b)$ represents the set of concrete elements $c_1 \cap c_2$. The reduction tries to find the smallest element $(a',b')$ such that the concretizations of $a'$ and $b'$ are subsets of those of $a,b$ (respectively), but their intersection remains the same as the original one ($c_1 \cap c_2$).

The lattice and semantic structures of the reduced product are exactly the same as those of the Cartesian product. In addition, a reduction operator aimed at refining the information tracked by the two domains is introduced, and it is used after each lattice or semantic operator application. Formally, the reduction operator $\rho : \funzione{\cartesiandomain}{\cartesiandomain}$ is defined by $\rho(\ael{c})=\bigsqcap_\cartesiandomain \{ \ael{c}' \in \cartesiandomain : \gamma_{\cartesiandomain}(\ael{c}) \leq_{\cartesiandomain} \gamma_{\cartesiandomain}(\ael{c}') \}$. Nevertheless, such definition is not computable in general, and often one wants to have a relaxed version of this operator that is not expensive to compute. In general, a reduction operator has to satisfy the following two properties: (i) $\rho(\ael{c}) \leq_\cartesiandomain \ael{c}$ (the result of its application is a more precise abstract element); (ii) $\gamma(\rho(\ael{c})) = \gamma(\ael{c})$ (an abstract element and its reduction represent the same property).

Consider again the example of the product of the Interval and Parity domains. A simple reduction operator may increase by one the lower bound (or decrease by one the upper bound) of the interval if the bound does not respect the information tracked by the Parity domain (e.g., it is odd while the parity tracks that the value is even). This way, the reduction of $([2..4], \cel{O})$, $([2..3], \cel{O})$, and $([3..4], \cel{O})$ yields in all cases the abstract value $([3..3], \cel{O})$. Note that the reduction operator does not always obtain the minimal information. For instance, if we reduce $([1..1], \cel{E})$ (where $\cel{E}$ represents the even element of the Parity domain), we would obtain $(\bot_I, \cel{E})$ (where $\bot_I$ is the bottom element of the Intervals domain), that could be further reduced to $(\bot_I, \bot_P)$ (where $\bot_P$ is the bottom element of the Parity domain). Therefore, the reduction operator usually requires to compute a fixpoint \cite{MIT}. 
As two other examples, consider the reduced product of Interval and Congruences domains. Firstly, the reduction of the abstract value $([2..2],3)$ produces the abstract value $(\bot,3)$ which is not a minimal element. We need to iterate the reduction to obtain $(\bot,\bot)$. Secondly, the reduction of $([4..5],2)$ produces the abstract value $([4..4],2)$ which can be reduced again to $([4..4],4)$.

Observe that the widening operator on the reduced product cannot be derived \textquotedblleft for free\textquotedblright\ as refinement on the the widening operators of the components. As proved in \cite{ZanioliCortesi}, this is true only under the (quite strict) condition that $\forall a_1,a_2 \in A, \forall b_1,b_2 \in B, (a_1\nabla_A a_2, b_1 \nabla_B b_2)\in \rho(A \times B)$, where $\rho(A \times B)$ represents the elements of the reduced product.
This property does not often hold in practice. A far simpler (and naive) solution consists in applying the widening component-wise and refraining from reducing the result before feeding it back as left argument of the next iteration' s widening. This subsumes the above condition (as the reduction becomes idempotent on the iterates), but it also allows converging when the condition does not hold.


\subsubsection{Complexity}
In addition to the complexity of the Cartesian product, the reduced product requires to compute the reduction operator. Therefore, the complexity of an operator of the reduced product is the sum of the complexity of the operators defined on $\firstdomain$ and $\seconddomain$ and of the reduction operator. Since this operator may require computing a fixpoint, the final cost of a generic operator could be rather expensive. Therefore, usually it is more convenient to define a reduction operator that refines only partially the information tracked by the two domains \cite{LOG08}.

\subsubsection{Implementation}
The implementation of the reduction operator has to be specific for the domains we are refining. Therefore, while the Cartesian product was completely generic and automatic, the reduced product requires one to define and implement how two domains let the information flow among them. This means that each time we want to combine two domains in a reduced product we have to implement such operator. On the other hand, all the other lattice and semantic operators are defined exactly as in the Cartesian product, except that they have to call the reduction operator at the end, but this can be implemented generically w.r.t. the combined domains.

\subsubsection{Granger product}
Granger \cite{G92} proposed an elegant solution to compute an approximation of the reduction operator. Granger based his new product on the definition of two operators $\rho_1 : \funzione{\cartesiandomain}{\firstdomain}$ and $\rho_2 : \funzione{\cartesiandomain}{\seconddomain}$. The idea is that each operator refines one of the two domains involved in the product. The final reduction is obtained by iteratively applying $\rho_1$ and $\rho_2$. In order to have a sound reduction operator, $\rho_1$ and $\rho_2$ have to satisfy the following conditions:
\begin{itemize}
\item $\rho_1(\ael{a}, \ael{b}) \leq_\firstdomain \ael{a} \wedge \gamma_\cartesiandomain(\rho_1(\ael{a}, \ael{b}), \ael{b}) = \gamma_\cartesiandomain(\ael{a},\ael{b})$
\item 
$\rho_2(\ael{a}, \ael{b}) \leq_\seconddomain \ael{b} \wedge \gamma_\cartesiandomain(\ael{a}, \rho_2(\ael{a}, \ael{b})) = \gamma_\cartesiandomain(\ael{a},\ael{b})$
\end{itemize}
The intuition behind Granger's product is that dealing with only one flow of information at a time is simpler. Each of the two operators $\rho_1, \rho_2$ tries to descend in one of the lattices: given the abstract element made by the pair $(\ael{a}, \ael{b})$, the $\rho_1$ operator uses the information from $\ael{b}$ to go down the lattice of 	$\firstdomain$, while the $\rho_2$ operator uses the information from $\ael{a}$ to go down the lattice of $\seconddomain$. After each application of $\rho_1$ or $\rho_2$ we get a smaller element. The descent is iteratively repeated until the operators cannot recover any more precision: the reduction operator $\rho(\ael{a}, \ael{b})$ is then defined as the fixpoint of the decreasing iteration sequence obtained by applying $\rho_1$ and $\rho_2$. This is defined by the sequence $(\ael{a}^n,\ael{b}^n)_{n \in \naturals}$ as follows:
\begin{align*}
(\ael{a}^0, \ael{b}^0) & = (\ael{a}, \ael{b}) \\
(\ael{a}^{n+1}, \ael{b}^{n+1}) & = (\rho_1(\ael{a}^n, \ael{b}^n),\rho_2(\ael{a}^n, \ael{b}^n))
\end{align*}

The Granger product has exactly the same complexity we discussed for the reduced product. The main practical advantage of the Granger product is that one only needs to define and implement $\rho_1$ and $\rho_2$, that is, how the information flows from one domain to the other in one step. Then the reduction operator relying on the fixpoint computation comes for free. 


\subsubsection{Open product}

Cortesi et al. \cite{CCH00} proposed a further refinement of the Cartesian product. Its purpose is to let the domains interact with each other during \emph{and} after operations by making explicit the domains' interaction through (abstract) queries. The open product is orthogonal to Granger's product and the two proposals can be combined, by incorporating Granger's idea of refinement inside the open product. The open product is orthogonal also to other methods such as down-set completion, and tensor product.

\subsection{Reduced cardinal power}
The reduced cardinal power was introduced by Cousot and Cousot in \cite{CC79}, but the literature concerning it has been relatively poor on both the theoretical and the practical level. The main feature of the cardinal power is that it allows one to track disjunctive information over the abstract values of the analysis. For instance, given the Interval and the Parity domain, one could track information like \textquotedblleft when \statement{x} is odd, \statement{y} is in [0..10]\textquotedblright. Some examples of the application of the cardinal power are the example 10.2.0.2 of \cite{CC79}, and examples 3 and 4 of \cite{CCL11}. In addition, a detailed explanation with various examples has been proposed by Giacobazzi and Ranzato \cite{GR99}. Let us look at the example in \cite{CCL11}, where the following slice of
code is analyzed (typical of data transfer protocols where even and odd numbered packets contain data of different types):
\lstset{numbers=none}
\begin{lstlisting}
1: n := 10; i := 0; A := new int[n];
2: while (i < n) do {
3:  A[i] := 0;
4:  i := i + 1;
5:  A[i] := -16;
6:  i := i + 1;
7: }
\end{lstlisting}
To analyze it, the authors combine \textit{Parity} (where the lattice is made by the abstract elements $\bot, o, e, \top$) and \textit{Intervals}. The reduced cardinal power of Intervals by Parity tracks abstract properties of the form $(o \rightarrow i_o, e \rightarrow i_e)$, which means that the interval associated to some variable is $i_o$ (resp. $i_e$) when the parity associated to another variable (which could be the same) is $o$ (resp. $e$). First of all, the authors show a non-relational analysis of the listing above, where they use:
\begin{itemize}
\item the reduced product of Parity and Intervals for simple variables;
\item the reduced cardinal power of Parity by Interval for array elements (hence ignoring their relationship to indexes)
\end{itemize}
For example $(o \rightarrow \bot, e \rightarrow [-16,0])$ means that the indexed array elements must be even with value included between $-16$ and $0$. The result of this analysis is: $i: (e, [10,10])$ (variable \statement{i} is even and has value $10$), $n: (e, [10,10])$ (variable $n$ is even and has value $10$), and $A: (o \rightarrow \bot , e \rightarrow [-16,0])$, which represents that array elements are abstracted by $(o \rightarrow \bot , e \rightarrow [-16,0])$ (i.e., they are even and with values in $[-16,0]$). The precision of this analysis can be greatly improved by using again the reduced cardinal power of Intervals by Parity, but this time relating the parity of an \textit{index} of the array to the interval of the \textit{elements} of the array at that index. The new result is $A: (o \rightarrow [-16,-16], e \rightarrow [0,0])$, which means that the array elements at odd indexes are equal to $-16$ while those at even indexes are $0$.

The reduced cardinal power has been formalized as follows in \cite{CC79}. Given two abstract domains $\firstdomain$ and $\seconddomain$, the cardinal power $\powerdomain=\seconddomain^{\firstdomain}$ with base $\seconddomain$ and exponent $\firstdomain$ is the set of all isotone maps $\powerdomain : \funzione{\firstdomain}{\seconddomain}$. Roughly, the combination of two abstract domains in $\seconddomain^{\firstdomain}$ means that a state in $\firstdomain$ implies the abstract state of $\seconddomain$ it is in relation with. The partial ordering $\leq_\powerdomain$ is defined by $\ael{f} \leq_\powerdomain \ael{g} \Leftrightarrow \forall \ael{x} \in \firstdomain : \ael{f}(\ael{x}) \leq_\seconddomain \ael{g}(\ael{x})$. Similarly, the least upper bound and greatest lower bound operators are defined as the pointwise application of the operators of $\seconddomain$. This way, $\langle \powerdomain, \leq_\powerdomain, \sqcup_\powerdomain, \sqcap_\powerdomain \rangle$ forms a lattice.

Let $f_1,f_2$ be two abstract elements in $\powerdomain$. Then, the widening operator $\nabla_{\powerdomain}$ on $(f_1,f_2)$ can be defined as:
$$\forall x \in \firstdomain : \nabla_{\powerdomain}(f_1,f_2)(x) = \nabla_{\seconddomain}(f_1(x),f_2(x))$$
Observe that the operator above can be effectively applied only if $\firstdomain$ is finite.

By defining $\alpha_\powerdomain(\cel{c}) = \lambda \ael{x} . \alpha_\seconddomain(\cel{c} \sqcap_\concretedomain \gamma_\firstdomain(\ael{x}))$ and $\gamma_\powerdomain(\ael{p})$ consequently, we have that $\langle \concretedomain, \leq_\concretedomain \rangle \galois{\alpha_\powerdomain}{\gamma_\powerdomain} \langle \powerdomain, \leq_\powerdomain \rangle$. We refer the interested reader to \cite{GR99} (and in particular to Theorem 3.6 and Proposition 3.7, where $\odot$ corresponds to $\sqcap_\concretedomain$) for more details and formal proofs.


A correctness result was presented in \cite{CC79} as well (Theorem 10.2.0.1). In this work, the authors focused on a collecting semantics defined by a lattice of assertions which is a Boolean algebra. Afterwards, Cousot and Cousot did not broaden their theoretical definition to a more general setting.

Let us recall the example used in \cite{CC79} to show the expressiveness of this domain.

\lstset{numbers=none}
\begin{lstlisting}
1: x := 100; b := true;
2: while b do {
3: 	x := x - 1;
4: 	b := (x > 0);
5: }
\end{lstlisting} 

The exponent of the cardinal power we use to analyze this example is the Boolean domain for variable $\statement{b}$, while the base is the Sign domain for variable $\statement{x}$ tracking values $\cel{+}, \cel{-}, \cel{0}$ as well as $\cel{0+}$ (meaning that the values are $\geq 0$), $\cel{0-}$ (meaning that the values are $\leq 0$), $\neq\cel{0}$ (meaning that the values are different from 0). This way, we track that when variable \statement{b} has a particular Boolean value, then the sign of variable \statement{x} has a particular sign. Before entering the \statement{while} loop, we know that $\statement{b}=\statement{true} \Rightarrow \statement{x}=\cel{+}$, while $\statement{b}=\statement{false} \Rightarrow \statement{x}=\bot$. After the application of the semantics of statement 3, we will have that $\statement{b}=\statement{true} \Rightarrow \statement{x}=\cel{0+}$, and $\statement{b}=\statement{false} \Rightarrow \statement{x}=\bot$, because the value of \statement{b} is unchanged (it is certainly true) while the value of \statement{x} has been decreased by one (so it could become equal to zero or remain greater than zero). After line 4, we obtain that $\statement{b}=\statement{true} \Rightarrow \statement{x}=\cel{+}$, and $\statement{b}=\statement{false} \Rightarrow \statement{x}=\cel{0}$, since the new condition \statement{x>0} is assigned to \statement{b}. In fact, \statement{b} equals to \statement{true} implies that \statement{x} must be greater than zero. In addition, we knew that the value of \statement{x} was $\geq 0$. Then, if \statement{b} is now \statement{false}, we are sure that \statement{x} will be equal to zero (but not less than zero). The fixpoint computation over the while loop stabilizes immediately (because, if we enter the loop again, we know that \statement{b} is true and, as a consequence, \statement{x} is positive, thus returning to the same conditions of the first iteration), and so we obtain that at the end of the program we have that $\statement{b}=\statement{true} \Rightarrow \statement{x}=\bot$, and $\statement{b}=\statement{false} \Rightarrow \statement{x}=\cel{0}$, since we have to assume the negation of \statement{b} to terminate the execution of the \statement{while} loop.

The cardinal power effectiveness is compromised when $\firstdomain$ is infinite (i.e., intervals in $\mathbb{Z}$), and can become costly when $\firstdomain$ is finite but non-trivial (i.e., intervals of machine integers). For this reason, some restricted forms of cardinal power can be used, where only a finite subset of $\firstdomain$ is represented.

Summarizing, the main difficulties in constructing a reduced cardinal power domain are: (i) the choice of elements in $\firstdomain$ to use; and (ii) the efficient design of abstract operators.

\subsubsection{Complexity}
Each time a lattice or semantic operator has to be applied to the abstract state, the cardinal power requires to apply it to all the elements of the base. Consider the cardinal power $\seconddomain^{\firstdomain}$. In a state of our domain, we will track a state of $\seconddomain$ for each possible state of $\firstdomain$. Let $n$ be the number of states of $\firstdomain$, $a$ the cost of an operator on $\firstdomain$ and $b$ the cost on $\seconddomain$. Then the overall cost over $\seconddomain^{\firstdomain}$ is $n*(a+b)$, since, for any element in $\firstdomain$ we have to apply the operator both on $\firstdomain$ and $\seconddomain$.

Let $h_a$ and $h_b$ be the height of the lattice of $\firstdomain$ and $\seconddomain$, respectively. Then the height of $\seconddomain^{\firstdomain}$ is $h_b^{h_a}$.

It is then clear that the cardinal power causes a significant increase in the complexity of the analysis w.r.t. the complexity of the two original analyses. If there is already practical evidence that the reduction operator in the reduced product may induce an analysis that is too complex \cite{LOG08}, it is even more important to carefully choose the two domains combined in a cardinal power. Nevertheless, particular instances of the cardinal power are already used to analyze industrial software, and in particular in ASTR\'EE \cite{BCC03}. ASTR\'EE exploits the Boolean relation domain, which applies the cardinal power using the values of some particular Boolean program variables as exponent. In this way, the analysis tracks precise disjunctive information w.r.t. these variables. In addition, ASTR\'EE contains trace partitioning \cite{MR05}. This can be seen as a cardinal power in which the exponent is a set of manually provided tokens on which the analysis tracks disjunctive information. There are various types of tokens the user can provide: particular abstract values of a variable, the begin of an \statement{if} statement, etc. It is proved that, in practice, if an expert user provides the \emph{right} tokens, the resulting analysis can be quite precise preserving its performances at the same time.

\subsubsection{Implementation}
The implementation can be rather simple when using a programming language providing functional constructs. In fact, the most part of the cardinal power (namely, elements of the domain, and lattice and semantic operators) can be defined as the functional point-wise application of the operators on the base and the exponent. Instead, the implementation of the cardinal power may be more verbose using an imperative programming language, but we do not expect it represents a significant challenge.

\subsubsection{Reduced relative power}
Giacobazzi and Ranzato generalized the reduced cardinal power \cite{GR99}. For this purpose, the authors introduce the operation of \emph{reduced relative power} on abstract domains. As it happened with the cardinal power, the reduced relative power is based on two domains $\aset{A}$ and $\aset{B}$ (respectively, the exponent and the base) and is defined in a general and standard abstract interpretation setting. Its formal definition is $\aset{A} \xrightarrow{\odot} \aset{B}$, where $\odot$ is a generic operator used to combine concrete denotations. It is called \textquotedblleft reduced \emph{relative} power\textquotedblright\ because it is parametric with respect to $\odot$. The operator $\odot$ should be thought of as a kind of combinator of concrete denotations: the glb is a typical example, but another less restrictive combinator could be needed for some non-trivial applications. An example comes from the field of logic program semantics. The reduced relative power can be used to systematically derive new declarative semantics for logic programs by composing the domains of interpretation of some well-known semantics. In this case a concrete domain of sets of program execution traces is endowed with an operator of trace-unfolding that does not behave like a meet-operation (in particular, it is not even commutative). For more details, see Section 7 of \cite{GR99}. The definition of the reduced relative power is as follows. $\aset{A} \xrightarrow{\odot} \aset{B}$ consists of all the monotone functions from $\aset{A}$ to $\aset{B}$ having the shape $\lambda \ael{x} . \alpha_B(\cel{d} \odot \gamma_A(\ael{x}))$, where: (i) $\cel{d}$ ranges over concrete values, (ii) $\gamma_A$ is the concretization function of $\aset{A}$ and (iii) $\alpha_B$ is the abstraction function of $\aset{B}$. These monotone functions establish a dependency between the values of $\aset{A}$ and $\aset{B}$, and for this reason are called \emph{dependencies}. Intuitively, a dependency encodes how the abstract domain $\aset{B}$ is able to represent the \textquotedblleft reaction\textquotedblright\ of the concrete value $d$ whenever it is combined via $\odot$ with an object described by $\aset{A}$.

\section{Examples}
\label{sect:motivatingexamples}
In this Section, we discuss the application of the Cartesian product, the reduced product, and the cardinal power to some examples dealing with arrays. This way, we show the main features and limits of each combination of domains. The two abstract domains we will combine are Intervals \cite{CC77} and a relational domain that tracks constraints of the form $\statement{x} < \statement{y} + c$.

\subsection{Cartesian product}
As a first example, we consider a quite standard program that initializes to 0 all the elements of a given array. 
	\lstset{numbers=none}
\begin{lstlisting}
for(i=0; i < arr.length; i++)
	arr[i] = 0;
\end{lstlisting}

We want to prove that the array accesses are safe, that is, $\statement{i} \geq 0$ and $\statement{i} < \statement{arr.length}$, in particular when we perform \statement{arr[i]=0}. 

If we run the analysis using only Intervals, we obtain that $\statement{i}=[0..+\infty]$. In fact, this domain cannot infer any information from the loop guard \statement{i<arr.length} since it does not have any upper bound for \statement{arr.length}. This result suffices to prove the first part of our property ($\statement{i} \geq 0$) but not the second part ($\statement{i} < \statement{arr.length}$).
If we run the analysis using only a relational domain, we obtain that $\statement{i} < \statement{arr.length}+0$ when we analyze the statement \statement{arr[i]=0} thanks to the loop guard. In this way, we can prove the second part of the property, but not the first part.

Therefore, the two domains alone cannot prove the property of interest, while the Cartesian product can. In fact, it runs the two analyses in parallel, and at the end 
it takes from both domains the most precise result they get regarding the property to verify. Combining the two results, the entire property is proved to hold.

\subsection{Reduced product}
Let us introduce a more complex example. It receives as input an integer variable \statement{l}, and it creates an array with one element if $\statement{l} \leq 0$, and of \statement{l} elements otherwise. Then it initializes the first \statement{l} elements to zero.

\begin{lstlisting}
if(l<=0)
	arr = new Int[1];
else 
	arr = new Int[l];
for(i=0; i < l; i++)
	arr[i] = 0;
\end{lstlisting}

As before, we want to prove that when we perform \statement{arr[i]=0} we have that $\statement{i} \geq 0$ and $\statement{i} < \statement{arr.length}$. In particular, the critical property is the second one, since the first one is already proved by Intervals as explained before.

If we analyze this example with the Cartesian product defined before, we obtain that (i) $(\{\statement{arr.length} \mapsto [1..1], \statement{l} \mapsto [-\infty..0]\}, \emptyset)$ in the \statement{then} branch, and (ii) $(\{\statement{arr.length} \mapsto [1..+\infty], \statement{l} \mapsto [1..+\infty]\}, \{\statement{arr.length} < \statement{l}+1, \statement{l} < \statement{arr.length}+1\})$ in the \statement{else} branch. Then, when we compute the upper bound of these two states, we obtain only $(\{\statement{arr.length} \mapsto [1..+\infty], \statement{l} \mapsto [-\infty..+\infty]\}, \emptyset)$. This leads to infer that $(\{\statement{arr.length} \mapsto [1..+\infty], \statement{l} \mapsto [-\infty..+\infty], \statement{i} \mapsto [0..+\infty], \}, \{\statement{i} < \statement{l}+0\})$ inside the loop, but this cannot prove that $\statement{i} < \statement{arr.length}$.

We now define a specific reduction operator that refines the information tracked by the relational domain with Intervals. In particular, if Intervals track that $\statement{x} \mapsto [a..b], \statement{y} \mapsto [c..d]$ and we have that $b \neq +\infty \wedge c \neq -\infty$, then in the relational domain we introduce the constraint $\statement{x} < \statement{y} + k$ where $k=b-c+1$. Thanks to this reduction operator, we infer that, in the \statement{then} branch, $(\{\statement{arr.length} \mapsto [1..1], \statement{l} \mapsto [-\infty..0]\}, \statement{l} < \statement{arr.length} + 0)$. Thank to this reduction, when we join the two abstract states after the \statement{if} statement we obtain that $(\{\statement{arr.length} \mapsto [1..+\infty], \statement{l} \mapsto [-\infty..+\infty]\}, \{\statement{l} < \statement{arr.length} + 1\})$. This leads to infer (when we analyze \statement{arr[i]=0}) that $(\{\statement{arr.length} \mapsto [1..+\infty], \statement{l} \mapsto [-\infty..+\infty], \statement{i} \mapsto [0..+\infty], \}, \{\statement{l} < \statement{arr.length} + 1, \statement{i} < \statement{l}+0\})$, and the information tracked by the relational domain proves that $\statement{i} < \statement{arr.length}$.

\subsection{Reduced cardinal power}
We slightly modify the previous example. In particular, we create an array of one element if $\statement{l \leq 2}$, and we initialize all the elements in the array from the third to the $\statement{(l-1)}$-th element.

\begin{lstlisting}
if(l <= 2)
  arr = new Int[1];
else 
  arr = new Int[l];
for(i = 3; i < l; i++)
  arr[i] = 0;
\end{lstlisting}

As in the previous example, the main challenge is to prove the second part of the property (that is, $\statement{i} < \statement{arr.length}$) when executing \statement{arr[i]=0}.

First of all, we show that the reduced product of Intervals and our relational domain is not in position to prove this property.
In the \statement{then} branch, the Interval domain tracks that $\statement{l} \in [-\infty .. 2]$ and $\statement{arr.length} \in [1 .. 1]$. This information yields the strict lower bound relationship $\statement{l} < \statement{arr.length} + 2$ through the reduction operator we previously introduced. The abstract state associated to the \statement{then} branch is $(\{ \statement{l} \rightarrow [-\infty .. 2], \statement{arr.length} \rightarrow [1 .. 1] \}, \{ \statement{l} < \statement{arr.length} + 2\})$, while in the \statement{else} branch we obtain $(\{ \statement{l} \rightarrow [3 .. +\infty], \statement{arr.length} \rightarrow [3 .. +\infty] \}, \{ \statement{arr.length} < \statement{l} + 1, \statement{l} < \statement{arr.length} + 1 \})$. When we compute the join between these two states, we obtain $( \{ \statement{l} \rightarrow [-\infty .. +\infty], \statement{arr.length} \rightarrow [1 .. +\infty] \}, \{ \statement{l} < \statement{arr.length} + 2 \})$. In fact, the join of the constraints $ \statement{l} < \statement{arr.length} + 2$ and $\statement{l} < \statement{arr.length} + 1$ results in $\statement{l} < \statement{arr.length} + 2$. Finally, inside the \statement{for} loop we know (from the Interval domain and its widening operator) that $\statement{i} \rightarrow [3 .. +\infty]$. Moreover, the loop guard implies that, when we perform \statement{arr[i]=0}, $\statement{i} < \statement{l}$ holds. From $\statement{i} < \statement{l}$ and $\statement{l} < \statement{arr.length} + 2$, we obtain that $\statement{i} < \statement{arr.length} + 1$, that is weaker than the property of interest $\statement{i} < \statement{arr.length}$.

Now consider the reduced cardinal power of Intervals on \statement{l} as exponent and our relational domain as base. The \statement{then} branch is associated to the abstract state $[-\infty .. 2] \Rightarrow \{\statement{l} < \statement{arr.length} + 2\}$, and the \statement{else} branch to $[3 .. +\infty] \Rightarrow \{ \statement{arr.length} < \statement{l} + 1, \statement{l} < \statement{arr.length} + 1 \}$. The join between these two states simply creates a new abstract state which contains both informations, that is, $[-\infty .. 2] \Rightarrow \{\statement{l} < \statement{arr.length} + 2\}$ and $[3 .. +\infty] \Rightarrow \{ \statement{arr.length} < \statement{l} + 1, \statement{l} < \statement{arr.length} + 1 \}$. When we enter the while loop, we have to consider the two cases separately:
\begin{itemize}
\item in the first case, $\statement{l} = [-\infty .. 2]$. Then, the loop guard $\statement{i} < \statement{l}$ is surely evaluated to false: the loop is not executed, so we do not need to verify the property about array accesses. Therefore, when $\statement{i} < \statement{l}$ holds, we have that $[-\infty .. 2] \Rightarrow \bot$. This way, when we analyze \statement{arr[i]=0}, we can discard this case;
\item in the second case, we have that $\statement{l} = [3 .. +\infty]$ and $\statement{i} < \statement{l}$. Then, from the abstract state obtained after the \statement{if} statements and assuming the loop guard, we know that $[3 .. +\infty] \Rightarrow \{ \statement{arr.length} < \statement{l} + 1, \statement{l} < \statement{arr.length} + 1, \statement{i} < \statement{l} \}$. By combining $\statement{l} < \statement{arr.length} + 1$ and $\statement{i} < \statement{l}$, we obtain $\statement{i} < \statement{l} \leq \statement{arr.length} \Rightarrow \statement{i} < \statement{arr.length}$, which is exactly the property we wanted to prove. 
\end{itemize}

\section{Conclusion}
In this survey, we presented various product operators in the abstract interpretation theory. 
For each product, we formalized its main components, we discussed its complexity and the implementation efforts required to implement it. We pointed out that, while the complexity of the Cartesian product does not cause any practical problem, the reduced product may affect the performances of the analysis if the reduction operator is too precise. In addition, the cardinal power leads to a lattice whose height is exponential w.r.t. the heights of the combined domains, and therefore this product requires the user to carefully and manually choose how to combine the two domains.
Finally we presented some examples that underline the expressiveness and the limit of the different products, showing in which scenarios a product is satisfactory or when we have to choose a more complex product.

\subsection*{Acknowledgments}
We are very indebted to Dave Schmidt. Both in his articles and especially in his presentations and discussions he has transmitted to us the passion in discovering new problems and solutions, the attention to clarity and understandability, and the strength of humility.

Work partially supported by the PRIN-Miur Project \textquotedblleft Security Horizons\textquotedblright .

\bibliographystyle{eptcs}
\bibliography{bibliografia}

\begin{thebibliography}{10}
\providecommand{\bibitemdeclare}[2]{}
\providecommand{\surnamestart}{}
\providecommand{\surnameend}{}
\providecommand{\urlprefix}{Available at }
\providecommand{\url}[1]{\texttt{#1}}
\providecommand{\href}[2]{\texttt{#2}}
\providecommand{\urlalt}[2]{\href{#1}{#2}}
\providecommand{\doi}[1]{doi:\urlalt{http://dx.doi.org/#1}{#1}}
\providecommand{\bibinfo}[2]{#2}

\bibitemdeclare{inproceedings}{BCC03}
\bibitem{BCC03}
\bibinfo{author}{B{.} \surnamestart Blanchet\surnameend}, \bibinfo{author}{P{.}
  \surnamestart Cousot\surnameend}, \bibinfo{author}{R{.} \surnamestart
  Cousot\surnameend}, \bibinfo{author}{J{.} \surnamestart Feret\surnameend},
  \bibinfo{author}{L{.} \surnamestart Mauborgne\surnameend},
  \bibinfo{author}{A{.} \surnamestart Min\'e\surnameend}, \bibinfo{author}{D{.}
  \surnamestart Monniaux\surnameend} \& \bibinfo{author}{X{.} \surnamestart
  Rival\surnameend} (\bibinfo{year}{2003}): \emph{\bibinfo{title}{A Static
  Analyzer for Large Safety-Critical Software}}.
\newblock In: {\sl \bibinfo{booktitle}{Proceedings of the ACM SIGPLAN 2003
  conference on Programming language design and implementation}},
  \bibinfo{series}{PLDI '03}, \bibinfo{publisher}{ACM}, \bibinfo{address}{New
  York, NY, USA}, pp. \bibinfo{pages}{196--207}, \doi{10.1145/781131.781153}.

\bibitemdeclare{article}{CCH00}
\bibitem{CCH00}
\bibinfo{author}{Agostino \surnamestart Cortesi\surnameend},
  \bibinfo{author}{Baudouin~Le \surnamestart Charlier\surnameend} \&
  \bibinfo{author}{Pascal~Van \surnamestart Hentenryck\surnameend}
  (\bibinfo{year}{2000}): \emph{\bibinfo{title}{Combinations of abstract
  domains for logic programming: open product and generic pattern
  construction}}.
\newblock {\sl \bibinfo{journal}{Science of Computer Programming}}
  \bibinfo{volume}{38}(\bibinfo{number}{1-3}), pp. \bibinfo{pages}{27--71},
  \doi{10.1016/S0167-6423(99)00045-3}.

\bibitemdeclare{article}{ZanioliCortesi}
\bibitem{ZanioliCortesi}
\bibinfo{author}{Agostino \surnamestart Cortesi\surnameend} \&
  \bibinfo{author}{Matteo \surnamestart Zanioli\surnameend}
  (\bibinfo{year}{2011}): \emph{\bibinfo{title}{Widening and narrowing
  operators for abstract interpretation}}.
\newblock {\sl \bibinfo{journal}{Computer Languages, Systems {\&} Structures}}
  \bibinfo{volume}{37}(\bibinfo{number}{1}), pp. \bibinfo{pages}{24--42},
  \doi{10.1016/j.cl.2010.09.001}.

\bibitemdeclare{}{MIT}
\bibitem{MIT}
\bibinfo{author}{P.~\surnamestart Cousot\surnameend}: \emph{\bibinfo{title}{MIT
  course 16.399: Abstract Interpretation}}.
\newblock \urlprefix\url{http://web.mit.edu/16.399/www/}.

\bibitemdeclare{inproceedings}{CC77}
\bibitem{CC77}
\bibinfo{author}{P.~\surnamestart Cousot\surnameend} \&
  \bibinfo{author}{R.~\surnamestart Cousot\surnameend} (\bibinfo{year}{1977}):
  \emph{\bibinfo{title}{Abstract interpretation: a unified lattice model for
  static analysis of programs by construction or approximation of fixpoints}}.
\newblock In: {\sl \bibinfo{booktitle}{Proceedings of the 4th ACM
  SIGACT-SIGPLAN symposium on Principles of programming languages}},
  \bibinfo{series}{POPL '77}, \bibinfo{publisher}{ACM}, \bibinfo{address}{New
  York, NY, USA}, pp. \bibinfo{pages}{238--252}, \doi{10.1145/512950.512973}.

\bibitemdeclare{inproceedings}{CC79}
\bibitem{CC79}
\bibinfo{author}{P.~\surnamestart Cousot\surnameend} \&
  \bibinfo{author}{R.~\surnamestart Cousot\surnameend} (\bibinfo{year}{1979}):
  \emph{\bibinfo{title}{Systematic design of program analysis frameworks}}.
\newblock In: {\sl \bibinfo{booktitle}{Proceedings of the 6th ACM
  SIGACT-SIGPLAN symposium on Principles of programming languages}},
  \bibinfo{series}{POPL '79}, \bibinfo{publisher}{ACM}, \bibinfo{address}{New
  York, NY, USA}, pp. \bibinfo{pages}{269--282}, \doi{10.1145/567752.567778}.

\bibitemdeclare{article}{CC92a}
\bibitem{CC92a}
\bibinfo{author}{P{.} \surnamestart Cousot\surnameend} \& \bibinfo{author}{R{.}
  \surnamestart Cousot\surnameend} (\bibinfo{year}{1992}):
  \emph{\bibinfo{title}{Abstract Interpretation and Application to Logic
  Programs}}.
\newblock {\sl \bibinfo{journal}{Journal of Logic Programming}}
  \bibinfo{volume}{13}(\bibinfo{number}{2-3}), pp. \bibinfo{pages}{103--179},
  \doi{10.1016/0743-1066(92)90030-7}.

\bibitemdeclare{inproceedings}{CH78}
\bibitem{CH78}
\bibinfo{author}{P{.} \surnamestart Cousot\surnameend} \& \bibinfo{author}{N{.}
  \surnamestart Halbwachs\surnameend} (\bibinfo{year}{1978}):
  \emph{\bibinfo{title}{Automatic discovery of linear restraints among
  variables of a program}}.
\newblock In: {\sl \bibinfo{booktitle}{Proceedings of the 5th ACM
  SIGACT-SIGPLAN symposium on Principles of programming languages}},
  \bibinfo{series}{POPL '78}, \bibinfo{publisher}{ACM}, \bibinfo{address}{New
  York, NY, USA}, pp. \bibinfo{pages}{84--96}, \doi{10.1145/512760.512770}.

\bibitemdeclare{inproceedings}{CC94}
\bibitem{CC94}
\bibinfo{author}{Patrick \surnamestart Cousot\surnameend} \&
  \bibinfo{author}{Radhia \surnamestart Cousot\surnameend}
  (\bibinfo{year}{1994}): \emph{\bibinfo{title}{Higher-Order Abstract
  Interpretation (and Application to Comportment Analysis Generalizing
  Strictness, Termination, Projection and {PER} Analysis of Functional
  Languages), invited paper}}.
\newblock In: {\sl \bibinfo{booktitle}{Proceedings of the 1994 International
  Conference on Computer Languages}}, \bibinfo{publisher}{IEEE Computer Society
  Press, Los Alamitos, California}, \bibinfo{address}{Toulouse, France}, pp.
  \bibinfo{pages}{95--112}.

\bibitemdeclare{inproceedings}{CCFMMMR05}
\bibitem{CCFMMMR05}
\bibinfo{author}{Patrick \surnamestart Cousot\surnameend},
  \bibinfo{author}{Radhia \surnamestart Cousot\surnameend},
  \bibinfo{author}{J{\'e}r{\^o}me \surnamestart Feret\surnameend},
  \bibinfo{author}{Laurent \surnamestart Mauborgne\surnameend},
  \bibinfo{author}{Antoine \surnamestart Min{\'e}\surnameend},
  \bibinfo{author}{David \surnamestart Monniaux\surnameend} \&
  \bibinfo{author}{Xavier \surnamestart Rival\surnameend}
  (\bibinfo{year}{2005}): \emph{\bibinfo{title}{The ASTRE{\'E} Analyzer}}.
\newblock In: {\sl \bibinfo{booktitle}{Proceedings of the 14th European
  conference on Programming Languages and Systems}}, \bibinfo{series}{ESOP'05},
  \bibinfo{publisher}{Springer-Verlag}, \bibinfo{address}{Berlin, Heidelberg},
  pp. \bibinfo{pages}{21--30}, \doi{10.1007/978-3-540-31987-0\_3}.

\bibitemdeclare{inproceedings}{CCL11}
\bibitem{CCL11}
\bibinfo{author}{Patrick \surnamestart Cousot\surnameend},
  \bibinfo{author}{Radhia \surnamestart Cousot\surnameend} \&
  \bibinfo{author}{Francesco \surnamestart Logozzo\surnameend}
  (\bibinfo{year}{2011}): \emph{\bibinfo{title}{A parametric segmentation
  functor for fully automatic and scalable array content analysis}}.
\newblock In: {\sl \bibinfo{booktitle}{Proceedings of the 38th annual ACM
  SIGPLAN-SIGACT symposium on Principles of programming languages}},
  \bibinfo{publisher}{ACM}, \bibinfo{address}{New York, NY, USA}, pp.
  \bibinfo{pages}{105--118}, \doi{10.1145/1926385.1926399}.

\bibitemdeclare{article}{FGR96}
\bibitem{FGR96}
\bibinfo{author}{Gilberto \surnamestart Fil{\'e}\surnameend},
  \bibinfo{author}{Roberto \surnamestart Giacobazzi\surnameend} \&
  \bibinfo{author}{Francesco \surnamestart Ranzato\surnameend}
  (\bibinfo{year}{1996}): \emph{\bibinfo{title}{A Unifying View of Abstract
  Domain Design}}.
\newblock {\sl \bibinfo{journal}{ACM Computing Surveys (CSUR)}}
  \bibinfo{volume}{28}(\bibinfo{number}{2}), pp. \bibinfo{pages}{333--336},
  \doi{10.1145/234528.234742}.

\bibitemdeclare{article}{FR99}
\bibitem{FR99}
\bibinfo{author}{Gilberto \surnamestart Fil{\'e}\surnameend} \&
  \bibinfo{author}{Francesco \surnamestart Ranzato\surnameend}
  (\bibinfo{year}{1999}): \emph{\bibinfo{title}{The Powerset Operator on
  Abstract Interpretations}}.
\newblock {\sl \bibinfo{journal}{Theoretical Computer Science}}
  \bibinfo{volume}{222}(\bibinfo{number}{1-2}), pp. \bibinfo{pages}{77--111},
  \doi{10.1016/S0304-3975(98)00007-3}.

\bibitemdeclare{inproceedings}{GR97}
\bibitem{GR97}
\bibinfo{author}{Roberto \surnamestart Giacobazzi\surnameend} \&
  \bibinfo{author}{Francesco \surnamestart Ranzato\surnameend}
  (\bibinfo{year}{1997}): \emph{\bibinfo{title}{Refining and Compressing
  Abstract Domains}}.
\newblock In: {\sl \bibinfo{booktitle}{Proceedings of the 24th International
  Colloquium on Automata, Languages and Programming}}, \bibinfo{series}{ICALP
  '97}, \bibinfo{publisher}{Springer-Verlag}, \bibinfo{address}{London, UK,
  UK}, pp. \bibinfo{pages}{771--781}, \doi{10.1007/3-540-63165-8\_230}.

\bibitemdeclare{article}{GR98}
\bibitem{GR98}
\bibinfo{author}{Roberto \surnamestart Giacobazzi\surnameend} \&
  \bibinfo{author}{Francesco \surnamestart Ranzato\surnameend}
  (\bibinfo{year}{1998}): \emph{\bibinfo{title}{Optimal Domains for Disjunctive
  Abstract Intepretation}}.
\newblock {\sl \bibinfo{journal}{Science of Computer Programming - Special
  issue on the 6th European symposium on programming}}
  \bibinfo{volume}{32}(\bibinfo{number}{1-3}), pp. \bibinfo{pages}{177--210},
  \doi{10.1016/S0167-6423(97)00034-8}.

\bibitemdeclare{article}{GR99}
\bibitem{GR99}
\bibinfo{author}{Roberto \surnamestart Giacobazzi\surnameend} \&
  \bibinfo{author}{Francesco \surnamestart Ranzato\surnameend}
  (\bibinfo{year}{1999}): \emph{\bibinfo{title}{The Reduced Relative Power
  Operation on Abstract Domains}}.
\newblock {\sl \bibinfo{journal}{Theoretical Computer Science}}
  \bibinfo{volume}{216}(\bibinfo{number}{1-2}), pp. \bibinfo{pages}{159--211},
  \doi{10.1016/S0304-3975(98)00194-7}.

\bibitemdeclare{inproceedings}{G92}
\bibitem{G92}
\bibinfo{author}{Philippe \surnamestart Granger\surnameend}
  (\bibinfo{year}{1992}): \emph{\bibinfo{title}{Improving the Results of Static
  Analyses Programs by Local Decreasing Iteration}}.
\newblock In: {\sl \bibinfo{booktitle}{Proceedings of the 12th Conference on
  Foundations of Software Technology and Theoretical Computer Science}},
  \bibinfo{series}{LNCS}, \bibinfo{publisher}{Springer-Verlag},
  \bibinfo{address}{London, UK, UK}, pp. \bibinfo{pages}{68--79},
  \doi{10.1007/3-540-56287-7\_95}.

\bibitemdeclare{article}{J97}
\bibitem{J97}
\bibinfo{author}{Thomas~P. \surnamestart Jensen\surnameend}
  (\bibinfo{year}{1997}): \emph{\bibinfo{title}{Disjunctive Program Analysis
  for Algebraic Data Types}}.
\newblock {\sl \bibinfo{journal}{ACM Transactions on Programming Languages and
  Systems (TOPLAS)}} \bibinfo{volume}{19}(\bibinfo{number}{5}), pp.
  \bibinfo{pages}{751--803}, \doi{10.1145/265943.265966}.

\bibitemdeclare{inproceedings}{LOG08}
\bibitem{LOG08}
\bibinfo{author}{F.~\surnamestart Logozzo\surnameend} \&
  \bibinfo{author}{M.~\surnamestart F{\"a}hndrich\surnameend}
  (\bibinfo{year}{2008}): \emph{\bibinfo{title}{Pentagons: A weakly relational
  domain for the efficient validation of array accesses}}.
\newblock In: {\sl \bibinfo{booktitle}{Proceedings of the 2008 ACM symposium on
  Applied computing}}, \bibinfo{series}{SAC '08}, \bibinfo{publisher}{ACM},
  \bibinfo{address}{New York, NY, USA}, pp. \bibinfo{pages}{184--188},
  \doi{10.1145/1363686.1363736}.

\bibitemdeclare{inproceedings}{MR05}
\bibitem{MR05}
\bibinfo{author}{L.~\surnamestart Mauborgne\surnameend} \&
  \bibinfo{author}{X.~\surnamestart Rival\surnameend} (\bibinfo{year}{2005}):
  \emph{\bibinfo{title}{Trace Partitioning in Abstract Interpretation Based
  Static Analyzers}}.
\newblock In: {\sl \bibinfo{booktitle}{Proceedings of the 14th European
  conference on Programming Languages and Systems}}, \bibinfo{series}{ESOP'05},
  \bibinfo{publisher}{Springer-Verlag}, \bibinfo{address}{Berlin, Heidelberg},
  pp. \bibinfo{pages}{5--20}, \doi{10.1007/978-3-540-31987-0\_2}.

\end{thebibliography}

\end{document}